# Killing Science Fiction:
# Why Conscious States Cannot Be
# Copied or Repeated

Andrew Knight
aknight@alum.mit.edu

**Abstract**

Several philosophical problems arising from the physics of consciousness, including identity, duplication, teleportation, simulation, self-location, and the Boltzmann Brain problem, hinge on one of the most deeply held but unnecessary convictions of physicalism: the assumption that brain states and their corresponding conscious states can in principle be copied. In this paper I will argue against this assumption by attempting to prove the Unique History Theorem, which states, essentially, that conscious correlations to underlying quantum mechanical measurement events must increase with time and that every conscious state uniquely determines its history from an earlier conscious state. By assuming only that consciousness arises from an underlying physical state, I will argue that the physical evolution from a first physical state giving rise to a conscious state to a second physical state giving rise to a later conscious state is unique. Among the consequences of this theorem are that: consciousness is not algorithmic and a conscious state cannot be uploaded to or simulated by a digital computer; a conscious state cannot be copied by duplicating a brain or any other physical state; and a conscious state cannot be repeated or created *de novo*. These conclusions shed light on the physical nature of consciousness by rendering moot a variety of seemingly paradoxical philosophy and science fiction problems.

# 1. Introduction

The nature of consciousness is fertile ground for a multitude of troubling, if not fascinating, thought experiments. There's the duplication problem: Imagine we can teleport a traveler to another planet by creating "a precise duplicate of the traveler, together with all his memories, his intentions, his hopes, and his deepest feelings," but then we decide not to destroy the original copy? "Would his 'awareness' be in two places at once?" (Penrose, 1989, p. 27) There's the simulation problem: If conscious awareness can be uploaded onto a computer, then how do we know we aren't simulated minds in simulated universes? (See, e.g., Bostrom, 2003) In fact, because simulated universes are much less expensive, in terms of matter and energy, than actual universes, then if consciousness is indeed algorithmic, we are almost certainly one of vast numbers of simulated copies! There's the problem of self-location: if a psychopath tells you that he has created an exact physical copy of you and will torture it unless you pay a hefty ransom, then you should pay the ransom unless you are absolutely sure that *you* aren't the copy! (See, e.g., Elga, 2004) There's the problem of the Boltzmann Brain: If consciousness is just the result of atoms in a brain, what's to prevent a set of physically identical atoms, somewhere in the universe, from accidentally coming together in just the right way to create your brain? And what would that feel like?

Notice that each of these problems is a direct consequence of the *copiability* or *repeatability* of conscious states.[1] If it turns out, for whatever reason, that conscious states cannot actually be copied, then these problems disappear. Of course, we'd also have to have a good explanation for why they couldn't – and that's exactly what this paper aims to address.

Science fiction is full of fascinating plots that involve copying conscious states. Whether each of the above scenarios is actually possible is a scientifically empirical question, and given the rate of technology advance, it may be just a matter of time before each is tested. There are related problems that may never be empirically testable, such as, "Will a computer ever become conscious – and how would we know?" There is no consensus on how to measure the existence or level of consciousness in an entity. Some say that consciousness depends on the ability to pass a hypothetical "Turing test." Some say it depends on the level of complexity in neural networks. Some say it depends on certain activity in the brain. So how can we possibly learn anything about the nature of consciousness if it depends on a definition?

What I want to discover is whether a conscious state is copiable or not. It's easy to get bogged down in the meaning of "conscious" and lose sight of the fundamental empirical issue. It's as if the question, "Will a ball near Earth experience a force of gravity?" has been preempted by, "What do you mean by 'ball'?" Fine – let me ask a different question: "If I am near Earth, will *I* experience a force of gravity?" I am not interested in philosophical or semantic debates about the meaning of "conscious" or "identity," nor do I need to decide on any particular definition of consciousness or identity to know that I am conscious and I am me. My goal in this paper is not to engage in idle philosophical chatter or to further dilute the ether with untestable claims. My goal is to derive, if possible, an objectively correct and falsifiable prediction to several questions, among them: will it ever be possible to teleport a copy of myself to another planet? Will it ever be possible to upload my conscious awareness onto a computer so that I can outlive my physical death? Will it ever be possible, perhaps in millions of years, for a collection of atoms somewhere in the universe to accidentally come together in just the right way to create my conscious awareness? Will it be possible for me to experience these things?[2] The answer is yes only if my conscious states are fundamentally such that they could be copied or repeated. Are they?

---

[1] Throughout this paper, I'll treat copiability and repeatability of conscious states as meaning the same.

Asking these questions does not threaten the scientific assumption of physicalism, which holds that consciousness is a purely material phenomenon – i.e., consciousness results from an underlying physical state or configuration of matter. Physicalism does *not* require the copiability of physical or conscious states, which is a separate assumption. Having said that, copiability seems to be a common feature of the physical universe, evidenced by functionally identical mass-produced consumer products and readily copied computer software and DNA code. After all, if consciousness were to arise from execution of a particular algorithm – the claim of proponents of Strong AI – then why wouldn't it arise from execution of an identical algorithm on a different computer? If consciousness were to arise from a particular configuration of matter – in a robot, for example – then why wouldn't it arise from a functionally identical configuration of matter? In other words, while physicalism does not imply the copiability of conscious states, it might seem like it does, even though the above problems of duplication, simulation, identity, etc., arise as a direct consequence of this assumption. In other words, despite the myriad of apparent paradoxes that arise, the underlying assumption that conscious states are copiable is rarely questioned within the realm of physicalism (but see Aaronson, 2016).

In this paper I will argue against physicalism's assumption of the copiability of conscious states. In Section 5, I will attempt to prove the Unique History Theorem – essentially, that every conscious state uniquely determines its history from an earlier conscious state – and in Section 6 I will discuss its implication that a conscious state cannot be copied or repeated. To do so, I must first discuss concepts like correlation, entanglement, and measurement in Section 2, the basic assumptions of the proof in Section 3, and a thought experiment exploring history dependence in conscious states in Section 4.

## 2.    Quantum Correlation and Measurement

It is natural, but incorrect, to regard an object as made of individual particles that can each be described independently of the object. Consider, for example, particles A and B, each of which can be in state |0> or |1>. Prior to interaction, the wave states of A and B can be written as a superposition of their possible states:

---

[2] Those readers who want to debate whether I am conscious or whether these questions have objectively correct answers will likely gain little from this article.

$$|\psi_A\rangle = a_1|0\rangle + a_2|1\rangle, \quad |\psi_B\rangle = b_1|0\rangle + b_2|1\rangle,$$

with coefficient (or "amplitude") $a_1$ such that the probability of measuring outcome 0 when measuring particle A in the basis {0,1} can be calculated as $|a_1|^2$, and so forth. However, when the particles interact, their wave states become entangled to produce a single wave state:

$$|\psi_{AB}\rangle = z_{00}|0\rangle_A|0\rangle_B + z_{01}|0\rangle_A|1\rangle_B + z_{10}|1\rangle_A|0\rangle_B + z_{11}|1\rangle_A|1\rangle_B$$

That is, interaction between the particles makes it impossible to describe either particle independently of the other. While the example above is trivial, a more interesting example of entanglement obtains when, for example, $z_{01} = z_{10} = 0$. In this case, measurements of the states of the particles in the same basis are guaranteed to yield identical outcomes – i.e., perfectly correlated outcomes – independently of the separation between the particles. That is, even if the particles are spacelike separated, the measurements of the states of the two particles will give the same outcome, an intriguing effect dubbed by Albert Einstein as "spooky action at a distance."[3]

In fact, any interaction between physical systems entangles them to thereby increase their correlations. Consider, for example, the effect of a quantum event on a measurement apparatus initially in state $|\psi\rangle_{ready}$, specifically the measurement in the vertical basis of the spin of an electron in a superposition of states $|\uparrow\rangle$ and $|\downarrow\rangle$. The following shows the evolution of the system during measurement:

$$|\psi\rangle_{ready}(c_1|\uparrow\rangle + c_2|\downarrow\rangle) \rightarrow c_1|\psi\rangle_{up}|\uparrow\rangle + c_2|\psi\rangle_{down}|\downarrow\rangle,$$

where $|\psi\rangle_{up}$ is the state of the measuring apparatus that has measured and displayed "spin up," etc. Note that prior to measurement only the electron exists in a superposition state while afterward the measuring apparatus, too, exists in a superposition state. To be fair, the above evolution only shows the process of entanglement by which the output of the measuring device gets correlated to the corresponding state of the electron. It does not show "measurement" as is colloquially understood – i.e., it does not answer the questions of whether the spin was measured up or down, and according to whom? We were trying to measure the spin of an electron, a weird quantum mechanical object whose mathematical description includes simultaneous mutually exclusive possibilities. But instead of measuring an outcome, all we did was create an even bigger weird quantum mechanical object: a measuring device whose mathematical description includes simultaneous mutually exclusive possibilities.

---

[3] While the results will indeed be perfectly correlated, each measurement outcome will appear random to its local observer, thus preventing superluminal signaling via entanglement. Perfect correlation will only become apparent *after* the observers have had a chance to compare notes, a process limited by the speed of light. (See, e.g., Maudlin, 2011)

The obvious solution to this conundrum seems to be to send in an actual person to read the output on the measuring device. However, adding a human observer to the system requires that either we accept that the human herself goes into a superposition state of both witnessing an "up" measurement and witnessing a "down" measurement (e.g., the Many Worlds Interpretation of quantum mechanics), or the state nonlinearly "collapses" into either $|\psi>_{up}|\uparrow>$ or $|\psi>_{down}|\downarrow>$ (e.g., the Copenhagen Interpretation). Collapse interpretations are arguably more palatable to common sense, but they necessitate an explanation of how and when a measurement occurs. For instance, does measurement have something to do with the interaction between a "microscopic" system and a "macroscopic" system, as in GRW objective reduction? (See, e.g., Ghirardi *et al.*, 1986) Does it require consciousness?[4] Von Neumann (1932) famously showed that a quantum particle being measured gets entangled with the measuring device, which gets entangled with the table on which the measuring device sits, and so on, in a so-called von Neumann chain, until collapse of the overall wave state, and that the collapse may occur at any point along the chain until the subjective perception of the human observer. Nevertheless, because none of these interpretations of quantum mechanics is currently empirically distinct, their debate is purely philosophical (and, arguably, scientifically irrelevant). What is relevant, however, is the objective answer to this question: "If I measure the electron's spin with a measuring device, will I observe a definite outcome?" The answer is yes. No philosophical semantics need obfuscate this conclusion. By the time I've observed an outcome, a measurement has taken place.

Just as important as identifying what does constitute measurement is identifying what does not constitute measurement. For example, consider a double-slit experiment performed on a stream of particles aimed at a screen, the measuring device potentially capable of determining which slit each particle passes through. If no "which-path" information is collected, the particles hit the screen in a pattern consistent with the notion that they are waves that pass through both slits – that is, an interference pattern. If which-path information *is* collected, then the particles predictably act as bits of matter and the interference pattern disappears, almost as if the particles "know" whether or not they are being watched. However, amazingly, if the which-path information is completely and irretrievably erased, the interference pattern *reappears*, as if the particles had acted as waves all along. (See, e.g., Kim *et al.*, 2000) Even more oddly, the choice of whether or not to erase the which-path information can be made *after* the particles have already hit the screen – and, in principle, even millions of years later. In a so-called delayed-choice quantum eraser experiment, the choice of whether or not to erase which-path information appears to retroactively determine whether a particle is detected as having passed through one slit or both. (See, e.g., Aspect *et al.*, 1982) In other words, a decision made long after a double-slit experiment is performed can, it seems, retroactively determine whether particles act as waves or bits of matter.[5]

---

[4] While few physicists seriously entertain the notion that wave state reduction requires consciousness, de Barros and Oas (2017) showed that "any experiment trying to falsify [the consciousness-causes-collapse hypothesis] on the basis of its different dynamics is doomed."

[5] It is easy to confuse correlation with causation, particularly in the quantum world. Quantum eraser experiments show that there will be a correlation between a post-measurement decision by an observer and a pre-measurement

If we regard the interference pattern as having been created by particles whose path was not measured, then the quantum eraser experiments teach us that if all potential correlation information of a quantum measurement is completely and irretrievably erased, then the quantum measurement, on retrospect, did not occur. In other words, measurement, as a bare-bones requirement, must create a lasting physical correlation.

Therefore, this paper will adopt the conservative notion of measurement simply as lasting physical correlation, thus mooting the concern of whether and when a wave state collapse occurs. Having said that, the best evidence I have that a measurement has taken place is that I make a conscious observation that is correlated to the outcome of the measurement.

## 3. Conscious and Physical States

I make several assumptions in this analysis. First, I assume physicalism in the broadest sense: that the conscious state $C_1$ of a conscious entity results from, supervenes on, and is a function only of some underlying physical state $S_1$ of matter. Of course, there's no guarantee that a given physical state will create a conscious state – probably very, very few will – but if a conscious state exists then it depends entirely on the underlying physical state.

I also assume that copying physical state $S_1$ elsewhere in spacetime would produce the exact same conscious state $C_1$ of the exact same person. This seems obvious to me because if two identical physical states could produce conscious experiences that were distinct in *any* way (including conscious states experienced by different people), then that difference would have to reside in something non-physical, thus contradicting physicalism. Nevertheless, there is some debate in the philosophy literature about whether identical physical systems separated by space or time are "numerically" identical. But such esoteric objections are utterly irrelevant to the present analysis because the problems introduced in Section 1 conveniently disappear if identical copies did not produce identical people. After all, if copying a physical state $S_1$ elsewhere did not give rise to the same conscious state $C_1$ of the same person, then where's the problem? If, for example, creating an exact replica of my physical state on Mars (coupled with destroying the original me) did not have the effect of teleporting *me* to Mars, then I've answered the question at hand and do not wish to engage in further debate as to who the person created on Mars might be.

I also assume that a given conscious state might be created by more than one physical state. In other words, it may be the case that conscious state $C_1$ arises from any member of some large set $S_1^*$ of underlying physical states, and that creating any of these physical states will produce the same conscious state $C_1$ of the same person. For instance, if it were ever possible to upload my consciousness onto a computer so that I can outlive my physical death, then it must be that my

---

action by particles. Strange as this may be, it is not quite correct to say that a decision in the future can cause an event in the past, because the effect can only be observed *after* the decision is made. Just as one cannot transmit information faster than light via quantum entanglement, information cannot be sent into the past by "peeking" at the screen before the decision is made in a delayed-choice quantum eraser experiment to see which decision will be made.

actual physical state would produce the same conscious state of the same person (me) as the very different physical state of a computer executing a software version of me. If a conscious state could not be copied to create the *same* person, then I can give up hope of immortality via computer simulation and won't further debate who the person created on the computer might be.

Why talk of conscious states at all? Because I know I'm in one. I don't claim to know what consciousness is, but I do claim to be conscious. Further, physicalism assumes that a given conscious state arises from an underlying physical state, but what is that? Classically, we might think of the physical state of a system in terms of the three-dimensional positions and momenta, at a given time, of all of its constituent particles. Such a system is, in principle, perfectly deterministic as well as time-reversible.

However, the classical viewpoint is inconsistent with quantum mechanics. For instance, it can't be said that any given particle in a system even *has* a position or momentum until measured. Worse, while its position can be measured to nearly arbitrary precision, quantum uncertainty guarantees a trade-off in the precision to which we can simultaneously measure its momentum. For that reason scientists often imagine a particle as a Gaussian "wave packet" whose spread in the position basis is inversely related to its spread in the momentum basis. To further complicate the issue and as broached in Section 2, the mathematical description of a particle can't be separated from that of other particles with which it is correlated. Therefore one current understanding of the physical state of an isolated system is an infinite-dimensional wave state consisting of a superposition of mutually exclusive "pure" states, each pure state described by a complex amplitude such that the probability of measuring the system in that pure state is the square of the absolute value of its amplitude. In principle, this wave state is deterministic and time-reversible – unless and until we admit that we sometimes observe definite outcomes to measurements. In other words, it is neither. Further, because there is no way, even in principle, to determine the wave state's initial or current conditions, the quantum mechanical description of a physical system larger than a few atoms does no better a job at elucidating the concept of "physical state" than the classical description.

So when I assert that my conscious state $C_1$ is created by physical state $S_1$, I don't necessarily know what kinds of information or sorts of physical attributes specify that physical state $S_1$. I also don't know how big state $S_1$ is. Is it the local physical state of certain neural connections in my brain? The state of my entire brain? My body? The planet? The universe? In other words, I know I am in conscious state $C_1$, and I know that that state results from some underlying physical state $S_1$. But I don't claim to know much more.

That said, one thing I *can* claim about any isolated physical system, based on the analysis in Section 2, is that its state reflects quantum measurements. Because a quantum measurement causes a lasting physical correlation, then the physical state of an isolated system must embed its history of quantum measurements. Consider the effect of a quantum measurement "branching" event on a system in initial state $S_1$ where measurement "A" would result in a physical state $S_a$ that is correlated to outcome A while measurement "B" would result in a physical state $S_b$ that is correlated to outcome B. These two hypothetical physical states ($S_a$ and $S_b$) would have to evolve along forever distinct branches so as to maintain their respective correlations; they could never evolve to the same physical state $S_2$. If they could, then state $S_2$, and all future physical states, would necessarily be uncorrelated to the measurement result, in which case the measurement event could not have happened. In other words, quantum correlations do not decrease; physical state is history-dependent.

Consequently, one cannot, even in principle, produce a copy of some physical state by measuring and replicating its physical attributes (position, etc.). If a physical state inherently embeds its unique history of quantum measurements, then there is no shortcut to creating or copying a physical state. Physical state $S_2$ encodes its unique evolution, including all quantum measurement outcomes, from earlier physical state $S_1$. The only way to create physical state $S_2$ is to produce a system of particles that has the correct correlation relationship among its particles, and the only way to do that is to start with physical state $S_1$ and to keep one's fingers crossed that it accidentally – via a unique series of random quantum measurement outcomes – evolves to state $S_2$. And of course the only way to create physical state $S_1$ is to create an earlier state $S_0$ that happens to evolve in just the right way... and so on back. In other words, every physical state is uniquely defined by its history and, quite simply, cannot be created *de novo*.

Therefore, my assumptions regarding physical and conscious states are minimal: specifically, that I am in a conscious state and that that conscious state results from an underlying physical state that uniquely determines its history.

I'll also assume that the only source of indeterminism in the universe is quantum measurement. In other words, I'll assume that a given physical state evolves deterministically and predictably, punctuated only by branching events caused by quantum measurement. Whether free will exists and the extent to which its existence affects the present arguments will be saved for a future analysis. Setting aside the question of free will, it may strike some as odd to treat *any* conscious experience as depending on a quantum measurement. How do I know that conscious experience can't be described adequately with classical modeling? How do I know that consciousness could even be sensitive to quantum events? The answer is simple: I am only interested in nondeterministic branching events, and the only known physical means is through quantum measurement. Such events need not happen within the brain, and in fact I have made no assumptions at all about the function of the brain or its relationship to consciousness. It is also indisputable that quantum events are regularly amplified in the real world so as to cause conscious correlation: consider the simple example of a person measuring (and consciously correlating to) a radioactive decay by hearing a "click" from a Geiger counter. So, absent free will, any and all conscious branching must ultimately be caused by quantum measurement.

Finally, I use the colloquial expression "stream of consciousness" for clarity only without assuming anything about the extent to which conscious experiences are continuous or discrete; and I use the word "person" to refer to any conscious entity, whether that entity is a human possessing "wetware," a conscious artificial intelligence running software, or any other conscious creature.

## 4. How to Win the Lottery

A person in conscious state $C_1$ has just purchased a lottery ticket with numbers 31-41-59-26-53-58. At time $t_1$, he sits down to watch the live lottery results on television, the six numbers drawn one at a time until future time $t_2$. Given that any nondeterminism in the lottery ball selection is due to quantum fluctuations, he knows that there are potentially billions of possible future conscious states in which he might exist at time $t_2$. However, the state that interests him most is one particular conscious state, which we will call $C_2$, in which he is holding the winning lottery ticket. Given that the imagined events naturally leading to state $C_2$ are possible but very unlikely, the gambler wants to guarantee the desired outcome by directly creating state $C_2$. Now let's assume – and this is the key assumption – that it is possible to independently create conscious state $C_2$ at time $t_1$. It should not matter whether the person's physical matter is instantaneously reconfigured to produce conscious state $C_2$ or a different collection of matter is configured into an identical conscious state $C_2$, so for the sake of clarity, let us assume that the person opts for the latter solution. In other words, while the person exists at time $t_1$ in conscious state $C_1$ from physical matter $M_1$, conscious state $C_2$ is created at time $t_1$ from physical matter $M_2$. Because conscious state $C_2$ is now definitely a future state of conscious state $C_1$, the person in state $C_1$ can expect his conscious state to eventually become $C_2$.[6]

Note that the person does not care which stream of consciousness he actually experiences from time $t_1$ to $t_2$. He does not care if he actually observes the drawing of numbers 20-20-20-20-20-20 or even if, after the drawing of the third number, lightning strikes his television. All he cares about is that at time $t_2$, he witnesses on his television screen the winning lottery numbers of 31-41-59-26-53-58. So the question now is whether conscious state $C_2$ can only be reached by a single, unique stream of consciousness or whether there are others. In Fig. 1, three possible future conscious states $C_2$, $C_2'$, and $C_2''$ are shown (among countless others), with the shown paths indicating, conceptually, possible streams of consciousness from conscious state $C_1$. First stream of consciousness $SOC_a$ is the person's experience of watching, one by one, the selection of numbers 31, then 41, then 59, then 26, then 53, and finally 58, while second stream of consciousness $SOC_b$ is the person's experience of something else; in both cases, however, it is posited that the person ends up in conscious state $C_2$.

We need no explanation for $SOC_a$, which seems to be the more natural path to conscious state $C_2$. At state $C_2$, having experienced path $SOC_a$, the person has exactly the experience and memories one would expect from having experienced path $SOC_a$ – i.e., from time $t_1$ to $t_2$, he experiences the drawing of each of the numbers on his lottery ticket and in state $C_2$ he experiences having won the lottery. On the other hand, $SOC_b$ would seem to require some kind of memory modification at time $t_2$: from $t_1$ to $t_2$ he has an experience inconsistent with the moment leading up to winning the lottery, but then at $C_2$ he experiences having won the lottery. I will now argue that $SOC_b$ (and any stream of consciousness besides $SOC_a$) is not possible, that a future conscious state is determined by a unique stream of consciousness.

---

[6] I will ignore the objection by proponents of the Many Worlds Interpretation that conscious state $C_2$ of matter $M_2$ might be a different "person" than the person of conscious state $C_1$ of matter $M_1$, or the philosophical objection that identity wouldn't necessarily flow from state $C_1$ to $C_2$, and so forth. The present thought experiment hinges on the assumption that it is possible for the person to directly create the conscious state in which *he* has won the lottery, not some hypothetical doppelganger in a one-in-a-billion branch. If conscious state $C_2$ is not a future state of *him*, then why go to the trouble of creating state $C_2$?

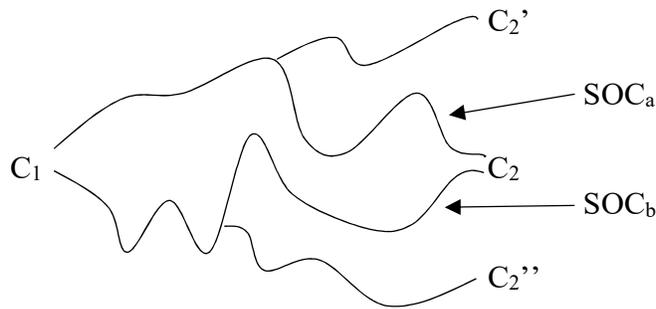

**Fig. 1. Different streams of consciousness evolving to the same conscious state.**

We have assumed that conscious state $C_2$ can be created *de novo* from matter $M_2$. Because state $C_2$ has literally just been created at time $t_1$, state $C_2$ is path-independent and has no history. In other words, even though state $C_2$ may include a consistent memory as if he *did* experience either $SOC_a$ or $SOC_b$, it makes no sense to speak of state $C_2$, having been newly created from matter $M_2$, as having actually experienced $SOC_a$ or $SOC_b$. However, conscious state $C_1$ (made of matter $M_1$) *must* evolve to state $C_2$: if the person embodied by matter $M_1$ wasn't guaranteed to eventually experience conscious state $C_2$, then it contradicts the original assumption that it is possible to create conscious state $C_2$ at time $t_1$. So the person embodied by matter $M_1$ *does* experience a stream of consciousness from time $t_1$ to $t_2$ – likely the natural stream $SOC_a$ instead of $SOC_b$, although it doesn't actually matter for the sake of argument. Thus the person in conscious state $C_2$ made of matter $M_1$ has, in fact, experienced a particular stream of consciousness, while the person in conscious state $C_2$ made of matter $M_2$ has not. But they are the *same* person! Conscious state $C_2$ must be the same whether state $C_2$ is created *de novo* or whether it evolves via $SOC_a$ or $SOC_b$, so the person in state $C_2$ (whether of matter $M_1$ or $M_2$) must have, in fact, experienced a particular stream of consciousness, which means that state $C_2$ can't be path-independent after all. Therefore, the only way to "create" a person in conscious state $C_2$ is to start with a person in conscious state $C_1$ and wait while he experiences the unique path ($SOC_a$) required to create state $C_2$. Said another way, if the person experiences state $C_2$, then he can be sure he's experienced the only stream of consciousness that led there from state $C_1$: $SOC_a$.

One objection to the above argument is that matter matters: it is not necessarily the case that a conscious state emerging from a configuration of matter $M_1$ is the same person as an identical conscious state emerging from a configuration of matter $M_2$. For instance, perhaps at time $t_1$ the person had his matter "rearranged" to produce conscious state $C_2$ so that he does not actually experience a stream of consciousness, unique or not, from state $C_1$ to $C_2$. Or perhaps at time $t_2$, when he is in some alternate conscious state $C_2'$ in which he believes he did not win the lottery, his matter is rearranged to effect a memory modification to state $C_2$, in which he believes he did win the lottery. These are merely quibbles. In fact the thought experiment was fatal from the outset because it assumed that a conscious state $C_2$, which depends on unpredictable quantum events that occur *after* time $t_1$, could be created at time $t_1$. The next section will more rigorously analyze why every conscious state uniquely determines its history, a conclusion that will be formalized in the Unique History Theorem.

## 5. Proof of the Unique History Theorem

Imagine some initial conscious state $C_1$ that is about to correlate to a quantum measurement event having possible outcome A or B. In other words, a quantum branching event is about to occur and the person's stream of consciousness will take either path $SOC_a$ that includes a conscious state $C_a$, correlated to measurement outcome A, or path $SOC_b$ that includes a conscious state $C_b$, correlated to measurement outcome B. My question is whether it is possible for streams $SOC_a$ and $SOC_b$ to evolve to the same conscious state $C_2$. If not, then conscious states, like their underlying physical states, are history-dependent. To answer this question, let's consider two scanarios:

*Scenario #1:* A person in some initial conscious state is about to do a quantum spin measurement on an electron, whose outcomes are A ("up") and B ("down"). The measurement apparatus is designed so that if the measurement outcome is A, then a large "A" flashes briefly on a screen; for B, a large "B" flashes. In the case of measurement A, the person experiences stream of consciousness $SOC_a$ that includes a conscious state $C_a$ in which she consciously observes seeing the letter "A" flash on the screen – i.e., state $C_a$ is correlated to measurement A. To be fair, in neither case has she *actually* observed the spin of an electron, which is far too small for human observation. What has really transpired is an amplification of a quantum measurement: consecutive correlated events amplified the initial measurement so that trillions of photons strike the screen in the shape of "A" or "B," the observed image correlated to the initial measurement. Still, even if the person hasn't directly observed the spin of the electron, she at least understands that her conscious state $C_a$, for example, is correlated to an "up" spin measurement outcome and is, in some sense, an observation of that outcome. However, correlation of a conscious state to a quantum measurement outcome need not be so blatantly related to observation of that outcome, as demonstrated by...

*Scenario #2:* Unbeknownst to some taxpayer, a random (quantum) error occurs in a computer used by the Internal Revenue Service such that if the error is measured as A, he will be mailed a refund check for $1000, but if it is measured as B, he will be mailed a letter demanding $1000. After receiving a sealed envelope from the IRS the following week, he will experience one of two competing streams of consciousness $SOC_a$ or $SOC_b$, each correlated to one of the possible measurement outcomes, even if he has no idea that the stream of consciousness he does experience is the result of some quantum measurement event, much less some measurement event in a remote IRS computer that occurred a week before.

Before further analyzing these scenarios, I should point out that a conscious correlation to a quantum measurement outcome might (at least at first) be barely detectable; perhaps it results in a slightly different aroma, a nuance of feeling, a mild change in taste. All that is required in this thought experiment is that the distinction *is* consciously detectable – i.e., that conscious state $C_a$, correlated to measurement A, is consciously distinct from state $C_b$, correlated to measurement B. So we might say that conscious state $C_a$ is correlated to measurement A if some aspect of conscious state $C_a$ is evidence that outcome A was measured, or perhaps that conscious state $C_a$ could not have been experienced unless outcome A was measured.

The question in both of the above scenarios is: could $SOC_a$ and $SOC_b$ evolve to the same conscious state $C_2$? Note that in order for both $SOC_a$ and $SOC_b$ to evolve to the *same* conscious state $C_2$, state $C_2$ must itself be uncorrelated to the quantum measurement event. So, in Scenario #1, for state $C_2$ to be uncorrelated to the measurement outcome, the experience of state $C_2$ cannot provide evidence that either "A" or "B" flashed on the screen. Consider the case in which conscious state $C_a$ is in fact experienced just before $C_2$; what would that feel like? What would it be like to consciously observe the letter "A" flash on a screen and then to experience a subsequent conscious state that is entirely uncorrelated to the quantum measurement event whose amplification resulted in that preceding conscious observation? Maybe it is like the experience of seeing the letter "A" flash on a screen and then instantly forgetting it.

However, Scenario #2 poses a greater conundrum. Imagine that state $C_a$ is the experience of first seeing a check made out to him for $1000 while $C_b$ is the experience of first seeing a demand for $1000 on IRS letterhead. For some future state $C_2$ to be uncorrelated to the measurement outcome, the experience of state $C_2$ cannot provide evidence of either the windfall (correlated to measurement A) or the debt (correlated to measurement B). The question is *not* whether there is some future conscious state $C_2$ in which he is not thinking about the outcome or has forgotten about the outcome; indeed we would expect that very few of his future conscious states would actively reflect on that windfall (or debt), and that there might even be a point in his future in which he has wholly forgotten about it. The question is whether it is possible for some future state $C_2$ to be entirely *uncorrelated* to the outcome. It is much harder to imagine such a state.

We can conceive of a series of consecutive conscious experiences starting at $C_a$ – for instance, he excitedly calls his mother, then jumps in his car to drive to the bank, then gets into a car accident that sends him to the hospital... and so forth. And we can conceive of a serious of consecutive conscious experiences starting at $C_b$ – for instance, he angrily kicks a hole in the wall, behind which he finds an antique container, inside which he discovers $1 million in gold bullion... and so forth. The question is whether it is conceivable that these two independent streams of consciousness could ever converge to the *same* conscious state, totally uncorrelated to the original measurement outcome? The answer seems intuitively "no" because states $C_a$ and $C_b$ were chosen to be so extreme and clearly distinct that it is hard to imagine any future stream of consciousness

that is not affected, in some way, by the experience of either $C_a$ or $C_b$. Every possible future conscious state seems to hinge on (and therefore embed the history of) the original outcome of A or B. Each possible stream $SOC_a$ and $SOC_b$ seems to consist of a series of conscious experiences, each experience correlated to its preceding experience, on backward in time to the original quantum measurement outcome of A or B. In other words, in each stream, conscious correlations grow such that every conscious experience he has afterward can be traced back to that original quantum glitch in the IRS's computer.

Based solely on the extreme example of Scenario #2, one might conclude that each possible stream of consciousness from a quantum branching event evolves chaotically and with ever increasing conscious correlations.[7] But is this *generally* true? What about when the conscious states $C_a$ and $C_b$ are not especially distinct, as in Scenario #1? For that observer, the differences between states $C_a$ and $C_b$ may be minimal. For example: she hasn't eaten all day and she's hungry; the temperature in the lab is too low and she's shivering; she's bored from doing so many spin measurements. Conscious states $C_a$ and $C_b$ may be identical in many respects, differing only in her observation of the letter that flashes on the screen. It is much easier to imagine the two similar (but distinct) conscious states of Scenario #1 evolving to an identical conscious state than the two extremely different conscious states of Scenario #2. Easy to imagine or not, is it actually possible for any two distinct conscious states to evolve to the same conscious state?

*More generally, is it possible for a conscious state that is correlated to a quantum measurement outcome to evolve to a conscious state that is not correlated to that outcome – i.e., is it possible for a conscious state to not uniquely determine its history?* To answer this question, I must consider the differences between physical and conscious states. As previously discussed, it is not possible for a physical state correlated to a quantum measurement outcome to evolve to a physical state uncorrelated to that same outcome, but conscious states are different in that there is no guarantee that conscious states always correlate to their underlying physical states. For instance, the physical state of a measuring device may correlate to the outcome of a quantum measurement event, but if an observer is not paying attention, his conscious state may not immediately correlate to the outcome.

Referring to Fig. 2, consider a tree of possible future physical states of state S, which creates a person's conscious state C. At some point, a quantum measurement event occurs so that S evolves either to $S_a$ (correlated to measurement A) or $S_b$ (correlated to measurement B). For whatever reason, this branch does not result in a correlated conscious state, so both $S_a$ and $S_b$ still map to conscious state C. This example assumes that conscious observation is not required to effect measurement, given that the corresponding conscious state does not change from S to $S_a$ (or $S_b$). It is assumed that measurement A causes (or is at least correlated to) a subsequent quantum measurement event whose outcome is either C or D. There is no reason to assume that measurement B causes the same subsequent quantum measurement event as measurement A, so it is assumed that the quantum event following outcome B will have either outcome E or F, different from C or D; and so forth.

---

[7] The 1998 film *Sliding Doors* popularized one version of this thought experiment. Imagine a person running toward the open doors of a subway car. A branching event occurs so that one possible physical state $S_a$ is such that the person is a little closer to the subway doors than that of another physical state $S_b$; conscious state $C_a$ of the person, created by underlying physical state $S_a$, may be nearly identical to conscious state $C_b$ created by state $S_b$, the main difference being in her observation of her distance from the subway doors. The subsequent closing of these subway doors serves as an amplification event whereby all of the protagonist's future conscious states can ultimately be traced back to the original branching event.

Fig. 2 is a highly simplified version of reality, intended to show only the possible relationships between physical states and their resulting conscious states. For example, deterministic time dependence is omitted: physical state S (and the conscious state C it creates) might evolve deterministically with time, but all I care about for the sake of this analysis are quantum branching events. Two additional comments about Fig. 2: first, while states $S_{ac}$ and $S_{be}$, for example, are shown in the same column, they need not occur at the same time; second, only binary branchings are shown for simplicity.

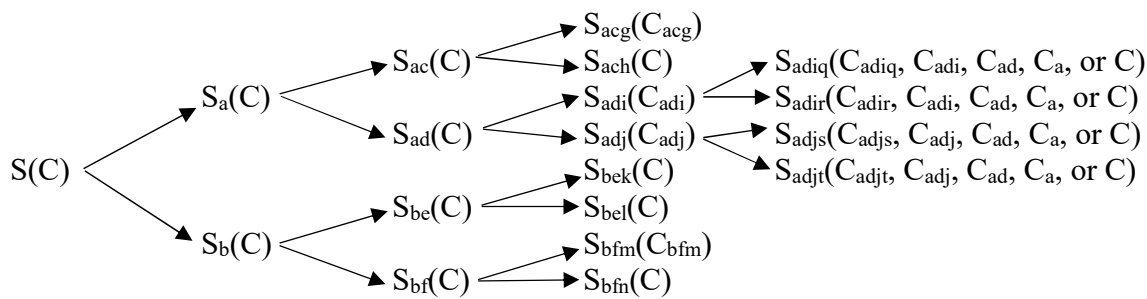

**Fig. 2. Tree of possible future physical and conscious states of physical state S.**

Note that because each measurement outcome makes possible the *next* measurement, each measurement outcome is also correlated with its every prior measurement outcome; this is the visual analog of the statement in Section 3 that every physical state uniquely determines its history.[8] For instance, assume that outcome G has been measured. Because the branching to G was made possible by outcome C, and the branching to C was made possible by outcome A, then outcome G correlates to outcomes C and A, as indicated below by $S_{acg}$. A similar nomenclature is used for conscious states. For instance, while physical state $S_{ad}$ is shown to produce the same conscious state C as both S and $S_a$ – in other words, outcomes A and D do not result in a correlated conscious state – the subsequent measurement outcome (either I or J) *is* correlated to a conscious state. (This could be explained simply as the person consciously noticing the measurement result.) But once a conscious state is correlated with an outcome (J, for instance), it is inherently correlated to every prior outcome – in this case A and D – for the same reason that underlying physical state $S_{adj}$ is so correlated. The person in conscious state $C_{adj}$ need not be consciously aware that A or D or even J have been measured; rather, state $C_{adj}$ is distinct from other possible conscious states in a way that depends on a measurement of outcome J (which depends on D, which depends on A, etc.).

---

[8] It might have seemed strange in Section 2 that quantum mechanics requires a measurement to create a "lasting" physical correlation. What does that mean? How long must we wait? Because measurement outcomes are correlated to past measurement outcomes, the conscious correlation of one measurement outcome ensures a lasting physical correlation of that outcome as well as every correlated outcome leading up to it.

Having explained the layout of the tree in Fig. 2, my goal is to discover if, and under what circumstances, a conscious state does *not* uniquely determine its history. I will attempt to find an example of a conscious state that can reach a possible future conscious state via two mutually exclusive conscious states.

Consider the conscious state (C) that arises from underlying state $S_{ad}$. What are the possible conscious states arising from $S_{adir}$ and $S_{adjs}$, for example, so that conscious state C (arising from $S_{ad}$) can reach a future conscious state (arising from either $S_{adir}$ or $S_{adjs}$) via either $C_{adi}$ (arising from $S_{adi}$) or $C_{adj}$ (arising from $S_{adj}$)? In Fig. 2, the possible conscious states of $S_{adir}$ and $S_{adjs}$ are shown – for instance, state $S_{adir}$ could produce any of conscious states $C_{adir}$, $C_{adi}$, $C_{ad}$, $C_a$, or C (or any prior conscious state). However, to ensure that both $C_{adi}$ and $C_{adj}$ could evolve to the *same* conscious state, then $S_{adir}$ and $S_{adjs}$ cannot produce a conscious state that is correlated to any event *after* measurement outcome D – that is, they must produce a conscious state of $C_{ad}$ or "less."

Let's analyze this particular example and add another column in Fig. 3. If this example was possible, it would demonstrate a counterexample to the assertion that every conscious state uniquely determines its history because the person's conscious correlations would *decrease* from state $C_{adi}$ (or $C_{adj}$). Notice in Fig. 3 that *future* conscious states of $C_{ad}$ *are* correlated to either outcome I or J, so the exception to the rule – the conscious state causing the logical problem – is $C_{ad}$.

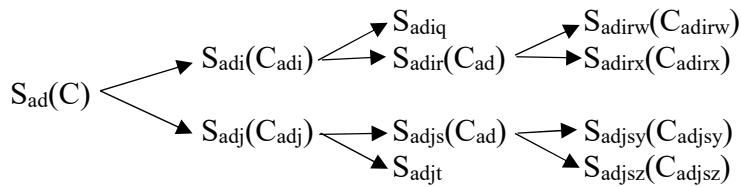

**Fig. 3. Tree showing conscious state C evolving to conscious state $C_{ad}$ via different streams of consciousness.**

Indeed it can easily be deduced that if it is possible that $C_1 \rightarrow C_2$ does not define a unique stream of consciousness, where $C_2$ is correlated to some recent measurement outcome M, then those non-unique streams of consciousness must include conscious states correlated to a measurement outcome *after* M. How could this be possible? And what would it feel like? In Fig. 3, what is the subjective experience of a person as his underlying physical state evolves from $S_{ad}$ to, say, $S_{adirx}$? Ostensibly, the stream appears as $C \rightarrow C_{adi} \rightarrow C_{ad} \rightarrow C_{adirx}$, but what does he actually experience?

One possibility is to accept the stream on its face: the person starts in state C, then experiences state $C_{adi}$ (perhaps by observing outcome I), then immediately forgets outcome I and experiences state $C_{ad}$ that is completely unrelated to outcome I, then experiences state $C_{adirx}$, which *is* correlated to outcome I (as well as R and X). I cannot imagine such an experience; perhaps such a subjective experience isn't possible.

Another possibility is that the person subjectively experiences a stream of consciousness that skips over the immediately forgotten state, i.e., C→$C_{ad}$→$C_{adirx}$. But if that's true, then Fig. 3 no longer represents a problem because conscious state $C_{ad}$ does then uniquely determine its history. Still another possibility is that the person subjectively experiences conscious states in order of increasing correlations, i.e., C→$C_{ad}$→$C_{adi}$→$C_{adirx}$. But again, if true, then Fig. 3 no longer represents a problem because conscious state $C_{ad}$ does uniquely determine its history.

It certainly seems that the example shown in Fig. 3 is either not possible or else is subjectively perceived in a way that conscious correlations do not decrease. Still, to understand what makes the example in Fig. 3 so strange, let me introduce the concept of *lag*. Lag represents the number of measurement outcomes to which an underlying physical state is correlated but the emergent conscious state is *not* correlated. Roughly, lag represents the number of times a physical state changes before a change is reflected in the emergent conscious state. Naturally, lag can increase by at most one in each step: either a physical state change gets reflected in a conscious state change or it doesn't. Consider, in Fig. 3, the path leading to state $S_{adirx}$:

| | |
|---|---|
| S produces C | lag 0 |
| $S_a$ produces C | lag 1 |
| $S_{ad}$ produces C | lag 2 |
| $S_{adi}$ produces $C_{adi}$ | lag 0 |
| $S_{adir}$ produces $C_{ad}$ | lag 2 |
| $S_{adirx}$ produces $C_{adirx}$ | lag 0 |

The apparent inconsistency inherent in state $C_{ad}$ is made explicit above: the increase in lag from 0 to 2 in a single physical step, producing state $C_{ad}$ *after* state $C_{adi}$, thus a reduction in conscious correlation. While any counterexample to the assertion that conscious states are history-dependent can be shown to increase lag unnaturally, I have not shown that an unnatural lag increase (i.e., reduction in conscious correlation) violates physics.[9] Nevertheless, even if an unnatural lag is physically possible, how problematic would it be? It depends on the lengths of lags – how long do they last? Conscious awareness seems to update very fast; for example, refresh rates in video footage of significantly less than 20 frames per second are consciously perceptible. Therefore it seems that typical lags of the kind that might concern us must be significantly less than a second.[10] Is this a threat to the assertion that conscious states are history-dependent and that conscious correlations do not decrease? Not if we care about a conscious state $C_2$ that is downstream from a past conscious state $C_1$ by days, weeks, or years – or anything significantly more than a second. In other words, to the extent that conscious correlations could ever decrease, such an oddity would affect the present analysis only to within a very short time period of state $C_2$.

---

[9] Then again, if physical state determines conscious state, then it determines lag, and we might properly ask why the universe would care *when* we observe something, given we have already assumed that consciousness is irrelevant to measurement. A reduction in conscious correlation probably does violate physics.

Having said that, there are other reasons to think that the evolutions shown in Fig. 3 are not physically possible. Consider that when $C_{ad}$ is experienced, it is assumed to be history-independent – that is, that there is no fact of the matter about whether the person previously experienced $C_{adi}$ or $C_{adj}$. But that's problematic because $C_{ad}$ is in fact created by one of two possible physical states ($S_{adir}$ and $S_{adjs}$), only one of which will evolve to a physical state that creates a conscious state correlated to, for instance, event J. In other words, there is no such state as $C_{ad}$ – rather, there is "a state $C_{ad}$ whose next conscious state will be correlated to event I" and "a state $C_{ad}$ whose next conscious state will be correlated to event J." So even though we have assumed the existence of a $C_{ad}$ that is history-independent, the only way for a person to experience $C_{ad}$ is to first experience either $C_{adi}$ or $C_{adj}$, a history that will be reflected in future conscious states of $C_{ad}$. This reasoning seems to imply that no such history-independent $C_{ad}$ could exist and, consequently, that every conscious state inevitably evidences its preceding conscious states.

Therefore, in deciding whether a conscious state uniquely determines its history from an earlier conscious state: for a very brief history, such as a second or two, the answer *might* be no; for a history of any significant length, the answer is yes. Therefore:

> **Unique History Theorem: For any time period in which lag is insignificant, a unique stream of consciousness maps a conscious state $C_1$ to a possible future state $C_2$; every conscious state uniquely determines its history from an earlier conscious state.**

## 6. Implications of the Unique History Theorem

If my analysis has been correct, then the conscious state of a person is defined by her history of experiences; and her conscious state $C_2$ is determined relative to her earlier conscious state $C_1$ by the unique stream of consciousness that she experienced. There are many implications of this assertion.

First, a conscious state can't be repeated. Imagine, for example, the repeat of conscious state $C_2$ via $C_1 \rightarrow C_2 \rightarrow C_3 \rightarrow C_2$. This violates the Unique History Theorem because $C_2$ does not define a unique history from $C_1$. Second, a conscious state can't be created *de novo*, nor can one be copied by creating an identical conscious state out of different matter, because such a state would be independent of history.[11]

---

[10] Scenario #2, for example, involved a hypothetical lag of a week, but by the time his conscious state correlated to the quantum outcome, the possibility of a future conscious state that did not correlate to that outcome seemed not credible. The real worry in Fig. 3 arises when lag is low, as in Scenario #1 and the example in Section 4.

Consequently, all the science fiction problems broached in Section 1 conveniently disappear. For instance, if teleportation will ever be possible, it can't be by copying a conscious state or its underlying physical state.[12] Perhaps more surprisingly, consciousness can't be simulated or uploaded to a digital computer; if it could, then nothing would prevent repeating a conscious state. Also, software is inherently history-independent: the same software could be produced by a human programmer as by a group of chimpanzees randomly pecking at a typewriter. So if conscious states are history-dependent, they can't be created by executing software. Further, any algorithm can be executed, in principle, by any general purpose digital computer and can therefore be repeated; this implies that consciousness can't be algorithmic. This in turn implies that a digital computer cannot be conscious and that the underlying premise of Strong AI is false.

What about the implications regarding the relationship between brain and consciousness? It is generally assumed that the brain *creates* consciousness. However, the Unique History Theorem implies that a Boltzmann Brain isn't possible – and a Boltzmann Brain is, by its nature, the very embodiment of the assertion that "brain causes consciousness." So if there is no way to copy my conscious state by creating another brain that is physically identical to mine to any physically possible degree of precision, then I'm not sure what it means to say, "My brain creates my consciousness."

Even if an external observer cannot distinguish the physical state of one brain from that of another, whatever conscious states arise from these physical states (if any) are history-dependent and will be able, from the subjective "inside," to distinguish themselves.[13] For instance, if my current conscious state arises from an underlying physical state, both the physical and conscious states are history-dependent. If one were to attempt to copy *me* by simply duplicating my brain, that brain would not have the requisite history, so in the unlikely event that it produced any conscious state, that conscious state would not be mine.

In fact, a person's conscious state must depend on entanglements among huge quantities of matter beyond his brain, a fact ignored in any attempted duplication of his brain. Consider, for example, the stream of consciousness that a carpenter would experience while building a house. The person starts, say, at conscious state $C_1$, and the particular stream of consciousness he experiences evolves to (and thus yields) a unique conscious state $C_2$. Assuming low lag – i.e., that his conscious state more or less correlates to his underlying physical state in real time – his stream of consciousness includes a variety of experiences that correlate to actual events in the world. For instance, he has a conscious visual experience of watching a backhoe level the foundation's dirt.

---

[11] I exclude the very *first* conscious state of a person, which of course has no history. I have no more explanation for my first conscious state than I or anyone else does for the first physical state of the universe.

[12] Aaronson suggests quantum no-cloning as the mechanism by which teleportation may be consistent with the non-copiability of conscious states (2016).

[13] The problem of subjective versus objective evidence of consciousness may be less perplexing after all: the only way to measure and specify a person's conscious state is to obtain information about its entire history, information to which only the person is (and can be) privy.

If we consider the popular account of consciousness, then this experience can simply be described as the capturing of video information via his eyes and the (albeit imperfect) storage of that information in his brain. Conscious experience arises then, per the popular account, from brain processes related to the capture, storage, access, and manipulation of this and other sensory information. And while such processes are admittedly not understood, proponents of the popular account insist that were we to duplicate the carpenter's brain, then an identical conscious experience would arise due to these brain processes.

The problem with the popular account is that sensation does not equal experience – i.e., the carpenter's conscious visual experience of watching a backhoe level the foundation's dirt cannot be duplicated by creating and storing video information. That particular conscious experience was correlated to actual events, which means that somewhere in the world (e.g., the construction site), there is in fact several tons of dirt that was moved. The only way to have the carpenter's particular conscious experiences is to actually create the history of correlations specified by the conscious state, which is to say that actual dirt must be moved, and the evidence must remain in the form of *its* correlations with other physical systems (like the planet). The amount of information necessary to specify the history of correlations due to this single experience is absolutely massive. The only way to experience conscious state $C_2$ is to *be* the conscious person who has, in fact, experienced the unique stream of consciousness necessary to produce state $C_2$.

The Unique History Theorem in no way questions the fundamental tenet of physicalism: that consciousness results entirely from the physical configuration of matter. However, it does underscore the extent to which popular interpretations of physicalism understate the significance and uniqueness of both physical and conscious states. Besides instantly eliminating many age-old problems of identity, duplication, etc., the Unique History Theorem has several other implications of which space currently limits elaboration. For example, because conscious states are history-dependent, several other science fiction plots are impossible: resetting a conscious state; erasing a memory by creating a new conscious state lacking the memory; modifying a memory by creating a new conscious state with a fake memory; and so forth.

## 7.    Conclusion

I first pointed out that several of the most perplexing problems of science fiction arise from an unnecessary assumption of physicalism, specifically that conscious states can be copied or repeated. I then analyzed possible conscious branching events in terms of underlying physical states and their correlations to quantum measurement events. I then attempted to prove the Unique History Theorem by showing that a conscious state that is correlated to a quantum measurement outcome cannot typically evolve to a conscious state that is not correlated to that outcome. The theorem – which states that, for any time period in which conscious lag is insignificant, every conscious state uniquely determines its history from an earlier conscious state – implies that conscious states cannot be copied or repeated (including by duplicating a brain) and that consciousness cannot be algorithmic.

These conclusions are not at odds with physicalism but do conflict with the generally accepted (albeit unproven) assumption that consciousness arises from the brain and is copiable in principle. While these conclusions may be viewed skeptically, they have the advantage of instantly putting to rest a plethora of problems regarding identity, physics, and consciousness. For instance, without the ability to create or duplicate a conscious state, many conundrums simply disappear: teleportation by copying is not possible; consciousness cannot be simulated; self-location is easy; and Boltzmann Brains cannot just pop into existence.

## Acknowledgements

I wish to thank Scott Aaronson, Paul Tappenden, Eric Brende, and Christopher Lind for their guidance and comments. This research did not receive any specific grant from funding agencies in the public, commercial, or not-for-profit sectors.